\documentclass[letterpaper,11pt,draftclsnofoot,onecolumn]{IEEEtran} %
\usepackage[margin=1in]{geometry}

\usepackage[dvipdfmx]{graphicx}
\usepackage{xcolor}
\usepackage{url}
\makeatletter
\let\NAT@parse\undefined
\makeatother
\usepackage[plainpages = false, colorlinks=true, linkcolor=black!30!blue, citecolor = green!40!black, urlcolor = blue, filecolor=black, pagebackref=false, hypertexnames=false,  pdfpagelabels ]{hyperref}
\usepackage{amsmath} 
\usepackage{amssymb} 
\usepackage{amsfonts}
\usepackage[mathscr]{eucal}
\usepackage{setspace}
\usepackage{parskip}
\usepackage{color}
\usepackage{multirow}
\usepackage[caption=false]{subfig}

\usepackage[compress]{cite} %

\usepackage{verbatim}
\makeatletter
\newcommand{\verbatimfont}[1]{\def\verbatim@font{#1}}%
\makeatother
\verbatimfont{\ttfamily\small}

\newcommand{\bi}{\begin{itemize}}\newcommand{\ei}{\end{itemize}}
\newcommand{\be}{\begin{equation}}\newcommand{\ee}{\end{equation}}
\newcommand{\bee}{\begin{enumerate}}\newcommand{\eee}{\end{enumerate}}
\newcommand{\bea}{\begin{eqnarray}}\newcommand{\eea}{\end{eqnarray}}
\newcommand{\beas}{\begin{eqnarray*}}\newcommand{\eeas}{\end{eqnarray*}}
\newcommand{\bc}{\begin{center}}\newcommand{\ec}{\end{center}}

\newcommand{\trasp}{\ensuremath{^{\intercal}}}

\newcommand{\bfu}{\ensuremath{\mathbf{u}}}

\newcommand{\xred}{\ensuremath{x^{\mathrm{grid}}}}
\newcommand{\red}{\ensuremath{^{\mathrm{grid}}}}
\newcommand{\excoal}{\ensuremath{^{\mathrm{coal}}}}
\newcommand{\xcoal}{\ensuremath{x^{\mathrm{coal}}}}

\newcommand{\setS}{\ensuremath{\mathcal{S}}}

\newcommand{\setX}{\ensuremath{\mathcal{X}}}
\newcommand{\setU}{\ensuremath{\mathcal{U}}}

\newcommand{\coal}{\ensuremath{\mathcal{C}}}
\newcommand{\setN}{\ensuremath{\mathcal{N}}}
\newcommand{\setM}{\ensuremath{\mathcal{M}}}

\newcommand{\setE}{\ensuremath{\mathcal{E}}}

\newcommand{\mmg}{\ensuremath{\mathcal{G}}}
\newcommand{\setSP}{\ensuremath{\mathcal{S}_{\mathscr{P}}}}
\newcommand{\coalstr}{\ensuremath{\mathscr{P}}}
\newcounter{subeqn} \renewcommand{\thesubeqn}{\theequation\alph{subeqn}}%
\newcommand{\subeqn}{%
  \refstepcounter{subeqn}
  \tag{\thesubeqn}
}
\title{Coalitional Control\\
{\Large Cooperative game theory and control}
}
\author{Filiberto Fele, Jos\'e M.~Maestre, and Eduardo F.~Camacho
\thanks{Financial support by the FP7-ICT project DYMASOS (ref. 611281), Spanish MINECO project COOPERA (DPI2013-46912) and Junta de Andaluc\'ia project ``Gesti\'on \'Optima de Edificios de Energ\'ia Cero'' (P11-TEP-8129) is gratefully acknowledged. }
\thanks{The authors are with the Department of Systems Engineering and Automation, University of Seville, 41092 Seville, Spain. Corresponding author: F.~Fele (\texttt{ffele@us.es}).}
\thanks{The final version of the article is available at \url{https://doi.org/10.1109/MCS.2016.2621465} (please cite as \cite{FELE_2017CSM}).}
\thanks{\copyright 2017 IEEE.  Personal use of this material is permitted.  Permission from IEEE must be obtained for all other uses, in any current or future media, including reprinting/republishing this material for advertising or promotional purposes, creating new collective works, for resale or redistribution to servers or lists, or reuse of any copyrighted component of this work in other works.}
}

\newif\ifPDF
\ifx \pdfoutput\undefined \PDFfalse 
\else
	\ifnum\pdfoutput > 0
		\PDFtrue 
	\else
		\PDFfalse
	\fi
\fi

\begin{document}
\maketitle

The evolution of information and communication technologies (ICT) has yielded means of sharing measures and other information in an efficient and flexible way~\cite{VadigepalliDoyle2003}, increasing the size and complexity of control applications~\cite{NEGENBORN10BOOK}. At the same time, the improvements in the computational and communicational capabilities of control devices have fostered the development of non-centralized control architectures, already motivated by the inherent structural constraints of large-scale systems. Computer-based control approaches such as model predictive control (MPC) are visible beneficiaries of these advances and have registered a significant growth regarding both theoretical and applied fields~\cite{SCA09JPC,NegenbornMaestreCSM2014}.\par
Whether or not the system arises from the interaction of different entities, it is generally possible to identify a set of coupled \emph{local} control problems, often defining a clear structure, that jointly configure the global one. Hence, the variables of a system can be grouped to highlight weakly coupled blocks: within each block (usually designated as \emph{neighborhood}) dynamic interactions propagate quickly, affecting the rest of the system on a longer time scale~\cite{StewartEtAl2010_ACC}.
In most cases centralized strategies do not exploit such structure, sometimes leading to unviable computational or communicational requirements.\par
As a natural way of avoiding the dependence on unavailable information---as well as to keep computational requirements at a minimum---it is desirable to formulate control laws based exclusively on local information~\cite{Siljak1991,ZecevicSiljak2010}. Non-centralized control strategies seek a tradeoff between performance loss and a scalable and flexible implementation~\cite{VadigepalliDoyle2003}. In general, though, the stronger the interaction among different parts of a system, the denser the communication required between the control agents---the extreme case corresponding to a distributed solution of the centralized control problem~\cite{RawlingsStewart2008}.\par
An extra degree of flexibility in the optimization of the computational and communicational requirements can be derived by the online identification of subsystems' interactions and the consequent real-time adjustment of the control law structure. This is the idea behind \emph{coalitional control}, a novel theory inspired by cooperative games, where the control strategy adapts to the varying coupling conditions between the controllers, promoting the formation of coalitions---clusters of controllers that cooperate in order to benefit from a jointly optimized control action. The adaptation of the controller topology may be the result of a top-down architecture, that is, imposed by a supervisor~\cite{MAESTRE13OCAM, FeleJPC, nunez2015time}, or of a bottom-up approach~\cite{FeleACC2015}, an autonomous coalition formation process occurring between the control agents. In all cases, the outcome will be a dynamically evolving coalitional structure of the overall controller.\par
Coalitional control focuses on the local interests that motivate the controllers to assemble, an aspect so far rarely contemplated in the distributed control literature. Indeed, although a well-defined global organizational objective may be present in large-scale infrastructures, it is not uncommon for the individual components to show interests that do not align with the global one~\cite{TsitsiklisThesis}. The smart grid and the intelligent transportation system are clear examples: a consistent research effort is being devoted to the issues associated with their management, typically involving different game-theoretic models in order to grasp the complex interaction phenomena produced by their heterogeneous user population~\cite{SaadEtAlSmartGrid2012,Malandrino2015b}.\par
This article presents the main concepts and challenges in coalitional control, and the links with cooperative network game theory. In the remainder, a path is traced from the issues related with model partitioning---a fundamental step in the design of distributed controllers---to the solutions proposed in the literature whose principles delineate the idea of coalitional control.
\section{Controlling large-scale systems: the path to coalitional control}
Consider a large-scale system described by the following discrete-time model:
\begin{equation}\label{eq_glob_sys}
\begin{split}
x^{+} & = f(x,u),\\
y & = g(x),
\end{split}
\end{equation}
where $x\in\mathbb{R}^n$ and $u\in\mathbb{R}^q$ are the global state and control input vectors, respectively, constrained in the sets $\setX$ and $\setU$, and $y\in\mathbb{R}^p$ is the vector gathering the outputs of the system.\par
A standard practice is to implement controllers on top of an already defined plant's structure---usually following structural constraints and fault tolerance requirements. This typically involves a prior partition of~\eqref{eq_glob_sys} into a set $\setN = \left\{1,\ldots,N\right\}$ of submodels, such that the essential dynamics characterizing the system are retained. 
Dynamic interactions typically occur between adjacent subsystems: coupling effects sensed beyond neighboring subsystems are seen through intermediate subsystems. Hence, the system's topology appears as a fundamental guideline for the reduction of the overall information exchange, which represents a critical aspect of distributed cooperative MPC schemes~\cite{RawlingsStewart2008}. It is clear how closely related the problems of system partitioning and distributed control are. Coalitional control encompasses both of them, allowing the \emph{dynamic} partitioning of the system into cooperative components.
\subsection{Model partitioning}
Model partitioning consists in the assignment of subsets of the global state and input variables to each of the agents involved in the control of a system, such that $x=\{x_i\}_{i\in\setN}\in\mathbb{R}^n$, $u=\{u_i\}_{i\in\setN}\in\mathbb{R}^q$. According to the classification provided by~\cite{LarssonSkogestad2000}, the system decomposition can be either \emph{horizontal} or \emph{hierarchical}. The first type relates to the physical structure of the system, while the second is based on the nature of the process, its characteristic time scales, and the control objectives. Hierarchical decompositions provide larger flexibility in shaping the controller to the heterogeneity of many large-scale systems---different sampling times, asynchronous operation of the different parts~\cite{Al-GherwiEtAl2011,StewartEtAl2010_ACC}. A general systematic methodology is hard to define due to the manifold nature (subsystems' interactions, time-scale, communication constraints, privacy concerns) of controlled systems, so that ad hoc approaches are frequently implemented.\par
Once the global model has been partitioned, each resulting subsystem $i\in\setN$ is assigned to a local control agent that has partial knowledge of the overall system, such that its behavior can be described by
\begin{equation}\label{eq_ind_ss}
\begin{split}
x_i^{+} & = f_{i}(x_i,u_i) + w_i,\\
y_i & = g_{i}(x_i) + \varpi_i,
\end{split}
\end{equation}
where $x_i\in\mathbb{R}^{n_i}$ and $u_i\in\mathbb{R}^{q_i}$ are respectively the local state and input vectors, constrained in the sets $\setX_i$ and $\setU_i$ respectively, and $y_i\in\mathbb{R}^{p_i}$ is the local output vector.
The vector $w_i\in\mathbb{R}^{n_i}$ represents the measurable state disturbances resulting from the coupling with other subsystems,
\begin{equation}
w_i = \sum_{j\in\setM_i}f_{ij}(x_j,u_j),
\label{eq_ind_disturb}
\end{equation}
where $\setM_i \triangleq \{j\in\setN\setminus\{i\}: \text{there exists } (x_j,u_j) \in \setX_j\times\setU_j\,|\, f_{ij}(x_j,u_j)\neq\mathbf{0}\}$ is referred to as the \emph{neighborhood} of subsystem $i$. Similarly, $\varpi_i\in\mathbb{R}^{p_i}$ expresses the measurable disturbances on the local output. Notice that neither the external state and inputs $x_j$ and $u_j$ nor their relation $f_{ij}$ with the local ones are known a priori by the local agent, eventually inducing some modeling error due to possibly neglected state--state, input--state, or input--output interactions.\par
In the search for the optimal structure of the control system, an important limiting factor comes from the available computation and communication resources. For instance, the excessive partitioning of the system may reduce the model size and hence the computational requirements at each node, yet at the expense of an increased communication load---which in turn limits the possibilities of a parallel implementation of the control algorithm. Similarly, while a small sampling time allows to obtain a model that reflects the actual structure of the system, it will result in high communication rates between nodes, as well as in a shorter time available for computation~\cite{VadigepalliDoyle2003}.\par
Besides, more often than not, large-scale systems are characterized by multi-scale dynamics---typically slower overall dynamics arising out of a group of subsystems with fast dynamics. With specific attention towards the strong relationship between model decomposition and sampling time, a structural analysis for decomposition is presented in~\cite{VadigepalliDoyle2003}, oriented to the implementation of a hierarchical multi-rate estimation and control architecture. In~\cite{JilgStur2013_ECC}, the coupling structure of the plant is analyzed prior to the design of feedback laws in order to decompose the model into hierarchically coupled clusters. A two-layer multi-rate control for such hierarchical decomposition is proposed in~\cite{JilgStur2013_IFAC}, where information is exchanged at each time step within clusters of strongly coupled subsystems, while a slower communication rate is required between different clusters.\par
The interaction among subsystems can be as well viewed as multiplicative output uncertainty. In~\cite{SamyudiaEtAl94}, the selection of best model partition within a given set of candidates is based on their associated (open loop) uncertainty bounds. The implications of system decomposition on robustness are investigated in~\cite{Al-GherwiEtAl2010}. The robustness of distributed MPC schemes characterized by different levels of knowledge of the non-local dynamics of the system is evaluated by means of an $H_{\infty}$ index. The formulation of a multi-objective MINLP based on this index allows to seek a tradeoff between global robust performance and the degree of connectivity among agents (translated in terms of local interaction model coverage). Given $N$ subsystems, the MINLP considers $N(N-1)$ binary variables to choose the structure of the controller over all possible connections among the subsystems. Simulation results show how controllers characterized by dense connectivity are more sensitive to errors in the interaction models. When these are affected by significant uncertainty, a fully decentralized structure can provide higher robustness.\par
Several other methods have been proposed in the literature. In~\cite{AskeEtAl2008}, the most relevant variables of the system are controlled by a central coordinator MPC, and a set of decentralized controllers complete the control action by responding to the inputs of the coordinator. The work of~\cite{MOTEE03ACC} addresses the assignment of actuators to a given set of controllers on the basis of two criteria, in order to achieve submodels of manageable size for MPC control. The first is an open-loop criterion based on maximizing the connectivity of the weighted graph representing the system, as an expression of the Hankel norm resulting by the controllability and observability Grammians. The second is a closed-loop criterion that aims at minimizing the performance degradation due to model partitioning, measured through the MPC's cost function. A method for the derivation of a distributed model is suggested in~\cite{StewartEtAl2010}, based on the Kalman canonical form of the linear state-space model for each input-output pair~\cite{AntsaklisMichel97}.\par
An exhaustive study of the issues related to model partitioning for the decentralized control of large-scale systems can be found in~\cite{Siljak1991}. One contribution of~\cite{Siljak1991} is that of showing how graph theory methods are well suited to gain an insight about the structural controllability and observability of large-scale systems, and employ these as a basis for model decomposition. Graph theory methods have been as well employed for the analysis of the impact of system topology on controllability and observability in~\cite{LiuEtAl2011,SummersLygeros2014}. 
Based on graph partitioning is the work of~\cite{OcampoBovoPuig2011}, where a decomposition method providing a set of non-overlapping subgraphs, with balanced number of vertices and minimal number of interconnecting edges, was developed. 
More recently, a threshold on the Shapley value~\cite{SHAP53AMS} of the potential communication links between the controllers was proposed in~\cite{muros2015application} as a criterion for partitioning.\par
\subsection{Cooperative control}
We assume in the remainder that the objective of each local controller is to drive the subsystem's state towards the origin of the state space.  The cost of subsystem $i$ at any given time step $k$ is expressed by $\ell_i(x_i,u_i)$, referred to as the stage cost. 
At time $k$, a control sequence is derived by the solution of the MPC problem
\begin{subequations}
\label{eq_ind_MPC}
\begin{align}
\min_{\bfu_i}\, & J_i = \sum_{t=0}^{N_p-1} \ell_i(x_i(t|k),u_i(t|k)) + V_i(x_i(N_p|k)) \label{eq_ind_MPC_cost} \\
& \mathrm{s.t.} \nonumber\\
& x_i(t+1|k) = f_{i}(x_i(t|k),u_i(t|k)) + \hat{w}_i(t|k),\label{eq_mod_MPC}\\
& h_i^{\mathrm{in}}(x_i(t|k),u_i(t|k)) \leq 0,\,t = 0,\ldots,N_p,\label{eq_restrin_MPC}\\
& h_i^{\mathrm{eq}}(x_i(t|k),u_i(t|k))  = 0,\,t = 0,\ldots,N_p-1,\label{eq_restreq_MPC}\\
& x_i(0|k) = x_i(k).\label{eq_init_constr}
\end{align}
\end{subequations}
The first element of the minimizer $\bfu_i^{\ast} \triangleq \left[u_i(0|k)^{\ast}, u_i(1|k)^{\ast},\ldots, u_i(N_p-1|k)^{\ast}\right]$ is applied as input to the subsystem, and the problem solved again at subsequent time steps in a receding horizon fashion.\par
Notice that, in absence of measures from the rest of the system, an estimate of the disturbance term~\eqref{eq_ind_disturb} is employed in the solution of~\eqref{eq_ind_MPC}. For this reason, decentralized schemes generally require either assuming a priori bounds on the coupling between subsystems, or considering worst-case interactions, resulting in a loss of performance~\cite{VadigepalliDoyle2003}. Communication between local controllers allows to achieve a visible enhancement in the overall control performance; however, it is difficult to synthesize a general result, due to the dependence on the particular scheme used~\cite{ALetAL10JPC,maestre2015comparison}. It is essential that the communication serves as a means to reach a shared consensus---an improvement over the decision that each agent would be able to make by relying merely on locally available information.
Once the model partitioning is given, the control by means of a cooperative scheme requires to answer fundamental questions. We analyze some of them in the following.
\paragraph{How communication improves performance?}
According to the degree of \emph{selfishness} of the objective function employed, two categories of controllers emerge in the distributed MPC literature: \emph{noncooperative}---where the agents pursue a local objective---and \emph{cooperative}---where the influence on the rest of the system, including how non-local objectives are affected, is accounted for in the choice of the inputs locally applied by each agent. More specifically, a global objective---consisting of a given (weighted) combination of individual objectives---is optimized over local inputs $\bfu_i$, $i\in\setN$,
\begin{equation}
\min_{\bfu_i} J = \sum_{t=0}^{N_p-1} \ell(x_i(t|k),u_i(t|k),x_{-i}(t|k),u_{-i}(t|k)) + V(x_i(N_p|k),x_{-i}(N_p|k)),
\label{eq_coop_MPC_cost}
\end{equation}
subject to~\eqref{eq_mod_MPC}--\eqref{eq_init_constr} (subscript $-i$ designates all subsystems $j\in\setN\setminus\{i\}$). Several techniques can be applied to obtain feedback over the values of nonlocal variables in~\eqref{eq_coop_MPC_cost}: typically, controllers update these values over intermediate iterations, by direct exchange or through Lagrangian multipliers.\\
The importance of \emph{cooperation} among MPC control agents is underlined in~\cite{RawlingsStewart2008}: the exchange of information among subsystems provided with mutual interaction models does not constitute a sufficient guarantee for closed-loop stability, due to the \emph{competition} arising from the pursuit of conflicting objectives. Similar conclusions are given by~\cite{TroddenRichards2009}, showing that increased cooperation not always translates to a gain in performance; indeed, in some cases it may even lead to a performance loss. This emphasizes the importance of the criteria upon which the cooperation is based. In case that the interests of local controllers diverge from the global benefit, a reward scheme may be implemented at the supervisory layer in order to bring these interests as much in line as possible~\cite{TsitsiklisThesis}.
\paragraph{Who should communicate with whom?}
The control agents communicate through a data network whose topology can be described by means of the undirected graph $\mathcal{G}=(\setN,\setE)$, where to each subsystem in $\setN$ is assigned a node. Dependence of the optimal topology $\setE\subseteq\setN\times\setN$ of the communication graph on the coupling between the local control problems is expected. The answer can be derived from a case-oriented evaluation of the performance deterioration due to the absence of communication between a given pair of local controllers. Control architectures for large-scale systems typically include a supervisory level, which may act as a coordinator to set the necessary information flows~\cite{TsitsiklisThesis}.\\
The sharing of information in a networked system can be represented through a \emph{knowledge graph}, where each node stands for the model of a given subsystem, and the edges indicate that the information about the models is available to the agents controlling the pointed nodes. The \emph{connectivity} and the \emph{degree distribution} of such graph can be interpreted in quantitative terms~\cite{LangbortDelvenne2010}.\\
In~\cite{TroddenRichards2009}, a graph representing the coupling between subsystems is updated at each time step by identifying the constraints that were active at the previous time step. Then, the cooperating sets are formed by those subsystems that are connected by a path in the graph. Further control-related studies, such as~\cite{Ishii14CSM, Maestre2015a}, focus on the PageRank index, a variant of the eigenvector centrality measure, used to quantify the relevance of nodes from a coalition formation viewpoint.
\paragraph{What information should be available to each local agent?}
The answer to this problem depends on the coupling source and the type of distributed control strategy that is implemented. In general,
the need for information exchange is inversely proportional to the coverage of the locally available information. 
In~\cite{SunEl-Farra2008}, the control is performed by means of a (global) linear feedback law. Continuous exchange of information among agents is avoided by providing dynamic models of the coupled subsystems to each agent, used to predict an estimate of the evolution of its neighbors' state when no new measure is communicated. In~\cite{FeleJPC} the effect of input coupling is viewed---on the grounds of the slow dynamics of the plant under study---as a constant disturbance along the prediction horizon, which is then estimated with a Kalman filter. Usual choices are the expected state and/or input sequences~\cite{JMM11JPC,TroddenRichards2013}, or auxiliary coordination variables such as prices~\cite{NEG09NHM} or sensitivities~\cite{Scheu2010ACC}. See also~\cite{MAE14DME} for further details regarding the information exchange in different schemes.\\
However, regardless of the knowledge of the global dynamics of the system that may be available to each agent, the closed-loop behavior can range from stable and almost optimal, to unstable Nash equilibria if conflicting objectives are pursued. This sensible difference in performance is likely to be detected when dealing with strongly coupled subsystems~\cite{RawlingsStewart2008}.
\paragraph{How to deal with constrained communications?}
The system dynamics impose limits on the time available to make agreements over cooperative decisions. Limitations in bandwidth or in energy consumption, as well as communication delays or packet losses concerning the data link infrastructure, need to be taken into account. Some algorithms are more flexible than others in this regard, providing superior robustness to delays or failures in communication. For example, the strategy presented in~\cite{StewartEtAl2010} admits the injection of suboptimal control actions in order to relax the heavy communicational requirement typical of iterative distributed schemes. Through a hierarchical design, this same strategy has been later extended to allow the asynchronous update among different neighborhoods~\cite{StewartEtAl2010_ACC}. Another example of algorithms showing such properties can be found in~\cite{TroddenRichards2013}, where the robust tube feedback formulation admits steps with no updates (by employing shifted input sequences).
\subsection{Dynamic neighborhoods and coalitional control}
The coupling among subsystems can be categorized over three classes. In presence of \emph{dynamical coupling}, a subsystem's behavior may be influenced by the value of the state and/or input of some other subsystems. Output coupling is a particular case of state coupling. This class of coupling can be described as in~\eqref{eq_ind_ss}--\eqref{eq_ind_disturb}. When the system is characterized by constraints involving multiple subsystems, the coupling is seen \emph{through the constraints}. This wide class includes constraints on input and/or state variables (such as physical limits of the plant or common-pool resources), and constraints not involving physical limits but related with the objective functions of the subsystems (for example, limited benefit within a free market with limited demand). Such problems may be analyzed under a zero-sum games perspective. Notice that constraint-coupled subsystems are not necessarily dynamically coupled. In this case, constraints~\eqref{eq_restrin_MPC}--\eqref{eq_init_constr} take the form:
\begin{equation}
h^{\mathrm{in}}(x,u) \leq 0,\quad h^{\mathrm{eq}}(x,u) = 0,
\label{eq_coupl_constr}
\end{equation}
where $h^{\mathrm{in}}$ and $h^{\mathrm{eq}}$ cannot be fully decomposed into independent equations $h^{\mathrm{in}}_i(x_i,u_i)$, $h^{\mathrm{eq}}_i(x_i,u_i)$, $i\in\setN$.\\ 
Lastly, coupling can appear \emph{through the objective function}: subsystems are part of a larger interacting environment and, therefore, it is likely that external variables exert some influence on their performance. These variables are usually related with the state of other subsystems, and often express economic indices. Notice that this category may overlap with the previous ones.\par
Broadly speaking, a \emph{neighborhood} designates a group of agents whose control problems show appreciable coupling that can be sensed within a limited time delay. 
Although the coupling structure may be fixed for a given system, in most cases the effect of coupling fluctuates. At this point, a natural question is \emph{what to do when coupling varies with time}? Is it reasonable to consider \emph{time-varying neighborhoods}?\par
Some distributed control schemes in the literature have already moved in this direction. The notion of \emph{cooperating sets} is employed in~\cite{TroddenRichards2009, TroddenRichards2013}. 
Within any given cooperating set, one subsystem at each time step locally computes optimal control actions for all the subsystems involved in the set, even if only the results relative to that subsystem are broadcast. The rationale behind computing such ``hypothetical'' non-local control inputs is to optimize the individual strategy considering what others may be able to achieve, so that cooperation is indirectly promoted among the agents by favoring the best plans for everyone within the cooperating set.
The composition of such sets is updated according to a graph representing the active coupling constraints. Interestingly, the results of~\cite{TroddenRichards2009} show that the optimal cooperating set is not necessarily restricted to \emph{directly} coupled subsystems.\par
In the distributed MPC scheme of~\cite{LopesDeLimaEtAl2015}, the cost incurred by each local controller is dynamically adjusted to fulfill minimum local requirements, on the basis of situational altruism criteria. 
On a similar line is the work of~\cite{NUNEZ13ICICAS}, where a flexible hierarchical MPC scheme is proposed for a hydro-power valley, where the priority of the agents in optimizing their control actions can be rearranged according to the different operational conditions.\par
The work of~\cite{FeleJPC} investigates the design of a hierarchical control scheme characterized by some flexibility over the employ of the data network through which the local control agents exchange information. The application on an irrigation canal is considered as a case study. The optimized design of a communication topology and its associated set of decentralized feedback control laws can be posed as a mixed-integer problem, but generally suffers from high computational complexity as the number of control nodes and the possible communication links between them grow. This issue is addressed through a two-layer greedy approach in~\cite{FeleJPC}, with the aim of optimizing the data links usage and decomposing the global MPC problem in small sized subproblems whenever possible. The goal of the supervisory layer is to find the best compromise between control performance and communication costs by actively modifying the network topology. The actions taken at the supervisory layer alter the control agents' knowledge of the complete system, and the set of agents with which they can communicate. Each group of linked subsystems constitutes a coalition, independently controlled based on a decentralized MPC scheme managed at the bottom layer. This feature is particularly interesting for communication infrastructures based on battery-powered wireless communication devices. The properties of a multi-agent control scheme based on this same idea are discussed in~\cite{MAESTRE13OCAM}, where the time-variant relevance of the communication within a set of dynamically coupled linear systems is analyzed using tools from cooperative game theory.\par
A different approach is found in~\cite{ValenciaThesis2012}, where the distributed MPC problem for dynamically-coupled subsystems is analyzed under a \emph{dynamic bargaining game} perspective. The formulation follows a few recent proposals in which the classic (static) bargaining theory have been extended to dynamic decision environments (see references in~\cite{ValenciaThesis2012}). These works focus on the \emph{coalition as the objective of the bargaining}, as a means of implementing a coalition-wide solution (the same for all members). However, the application of the same control input by all the agents in a coalition is likely to provide poor performances---if not an infeasible strategy---in a controlled system.
Filling this gap, the work of~\cite{ValenciaThesis2012} provides a distributed MPC formulation where cooperation is subject to bargaining: an agent accepts to be part of a coalition only if a benefit is foreseen over the performance expected by acting independently. Guarantees for the satisfaction of a minimum individual performance are imposed by means of a \emph{disagreement point}, defined as the threshold of maximum allowed loss of performance in case of cooperation. 
An agent's disagreement point is decreased whenever it decides to cooperate, and increased otherwise such as to foster a later participation; in this way, the disagreement point tends to the optimal expected value of the objective function.\par
The recent advances on ICT, particularly the spread of wireless networks and \emph{cloud-based} applications, simplify the deployment of a communication infrastructure between local controllers. The use of a distributed database of systemwide measures to achieve a globally optimal, yet scalable, control is a promising line of research yet to be explored~\cite{VadigepalliDoyle2003}. Nonetheless, issues derived by privacy-concerned subsystems---when non-local information is critical for adequate global feedback---not inclined to share local models and/or state information, need to be specifically addressed.\par
The distinct feature of such adaptive schemes, namely the rerouting of information flows among changing sets of controllers, can be schematized by the dynamical graph $\mmg(k) = (\setN,\setE(k))$, where the time dependence of $\setE(k)\subseteq \setN \times \setN$ reflects the possibility to activate or shut down data links at any given time step $k$. Let's now focus on the possibilities offered by the \emph{cooperation} among the controllers. A coalition is constituted by establishing flows of information---a broadened control feedback---within a given set of agents. The cooperation of the agents within a coalition can be carried out in different ways: the agents can exchange whichever information is required to enhance their performance by jointly solving the control problem---commonly state or output trajectories, or planned input sequences. The flow of such information from agent $i$ to agent $j$ is enabled by the activation of the associated link $\ell_{ij}=\{i,j\}\in\setE$.\par
The description provided by $\mmg(k)$ delineates a partition $\mathscr{P}(\setN,\mmg(k)) = \{\coal_1,\ldots,\coal_{N_c}\}$ of the set of controllers into $N_c$ connected components, referred to as coalitions~\cite{MSAVDN01}. Coalitions are disjoint sets such that~\cite{Rahwan2012}
\begin{equation*}
\coal_i \subseteq \setN,\, \text{for all } i\in\{1,\ldots,N_c\},\,\text{and } \bigcup_{i=1}^{N_c} \coal_i =\setN.
\end{equation*}
The number of coalitions $N_c$ pertains to the interval $\left[1,|\setN|\right]$, whose extremes correspond to the centralized control case (all the $|\setN|$ subsystems connected) and the case where each subsystem ``forms a coalition'' on its own (all links disabled).
The dynamics~\eqref{eq_ind_ss} of all subsystems relative to a given connected component $i\in\{1,\ldots,N_c\}$ can be aggregated as
\begin{equation}
\xi_i^{+} = F_{i}(\xi_i,\nu_i) + \omega_i,
\label{eq_coal_ss}
\end{equation}
with $\xi_i  \triangleq \{x_j\}_{j\in\coal_i}$, $\nu_i \triangleq \{u_j\}_{j\in\coal_i}$ the aggregate state and input vectors, and $F_{i}(\xi_i,\nu_i)$ the relative state transition function, describing the state and input coupling between members of the same coalition. Finally, the vector
\refstepcounter{equation}\label{eq_disturb_coal}
\begin{equation}
\omega_i = \{w_j'\}_{j\in\coal_i}\subeqn
\label{eq_disturb_coal1}
\end{equation}
gathers the disturbances due to the coupling with subsystems external to $\coal_i$. Following~\eqref{eq_ind_disturb} it holds that
\begin{equation}
w_j' = \sum_{r} f_{jr}(x_r,u_r),\text{with } r\in\setM_j\setminus\coal_i,\subeqn
\label{eq_disturb_coal2}
\end{equation}
pointing out how, for each $j\in\coal_i$, the set of unknown coupling from neighboring subsystems is reduced to the neighbors left out of the coalition. That is, from the coalition standpoint, the uncertainty comes from any subsystems $r\in(\bigcup_{j\in\coal_i}\setM_j)\setminus\coal_i$. Notice that in case of singleton coalition, i.e., $\coal_i \equiv \{i\}$, the description given by~\eqref{eq_coal_ss} coincides with~\eqref{eq_ind_ss}.
\subsection{Coalitional control objective}
In the remainder, we refer to the stage cost extended to $\coal_i\subseteq\setN$ as $\boldsymbol{\ell}_i(\xi_i, \nu_i)$ (in most cases $\boldsymbol{\ell}_i \equiv \sum_{j\in\coal_i} \ell_j$). Local controllers aggregate into a coalition with the aim of coordinating the effort and achieving a better overall performance. At time $k$, a control sequence for all subsystems $j\in\coal_i$ is derived by the joint solution of the MPC problem
\begin{subequations}
\label{eq_coal_MPC}
\begin{align}
\min_{\boldsymbol{\nu}_i}\, & \mathbf{J}_i = \sum_{t=0}^{N_p-1} \boldsymbol{\ell}_i(\xi_i(t|k),\nu_i(t|k)) + \mathbf{V}_i(\xi_i(N_p|k)) \label{eq_coal_MPC_cost} \\
& \mathrm{s.t.} \nonumber\\
&\xi_i(t+1|k)  = F_{i}(\xi_i(t|k), \nu_i(t|k)) + \hat{\omega}_i(t|k),\label{eq_coal_mod_MPC}\\
& H_i^{\mathrm{in}}(\xi_i(t|k),\nu(t|k)) \leq 0,\,t = 0,\ldots,N_p,\label{eq_coal_restrin_MPC}\\
& H_i^{\mathrm{eq}}(\xi(t|k),\nu(t|k))  = 0,\,t = 0,\ldots,N_p-1,\label{eq_coal_restreq_MPC}\\
& \xi_i(0|k) = \xi_i(k),\label{eq_coal_init_constr}
\end{align}
\end{subequations}
Problem~\eqref{eq_coal_MPC} is solved independently for each coalition $\coal_i\in\mathscr{P}(\setN,\mmg(k))$. Analogous to~\eqref{eq_ind_MPC}, at time $k$ the first element of $\boldsymbol{\nu}_i(k)^{\ast}$ is applied to every subsystem involved in the coalition.
Given~\eqref{eq_ind_disturb} and~\eqref{eq_disturb_coal2} we have $\left\|w_j'\right\|\leq\left\|w_j\right\|$ for $j\in\coal_i$; in words, since the effect of unmodeled interactions on the dynamics of any subsystem is reduced by its participation in a coalition, an improvement of the associated performance index is expected. Broadly speaking, provided the agents cost functions are linked, a prerequisite of cooperation is that the aggregate cost of the agents participating in a coalition outperforms the cost that the agents would have achieved through noncooperative optimization, that is, $\mathbf{J}_i^{\ast} < \sum_{j\in\coal_i} J_j^{\ast}$ must hold. In case the cost function expresses economical quantities, this translates in the availability of a surplus that can be split among the members of the coalition.\par
\subsection{Cost of cooperation}
Cooperation may not come for free. First, as previously discussed, communication may be constrained: a clear example is given by wireless networks, where the use of the links can be restricted in bandwidth/duration to reduce the energy consumption. In order to stimulate an optimal use of the network, it is reasonable to associate a cost to any communication (including those performed for the obtainment of measures). In some contexts, where privacy concerns are preponderant, the cost may also depend on the identities of the transmitter and the receiver: some agents may be less prone to exchange information than others. Besides, the coordination of a large number of agents can become a problem itself because the computation and communication requirements grow with the number of cooperating controllers involved. Therefore, costs required for the cooperation of a given set of agents can be taken into account by means of ad hoc indices related with the composition of the coalition or the data links needed in order to establish communication between every member of the coalition. Further measures may be employed to evaluate cooperation costs, based on, for example, the number of decision variables and/or constraints of the aggregate problem, reflecting the computational requirements. For coalition $\coal_i$, cooperation costs can be expressed as a function $\chi_i(\xi_i,\nu_i,\coal_i,\setE_{i}(k))$, where $\setE_{i}(k)\subseteq\setE(k)$ is the subset of edges of the graph $\mmg(k)$ connecting the nodes in $\coal_i$. Such cooperation costs can be assumed comparable with the stage cost. We can thus modify what stated at the beginning of the previous section in ``local controllers aggregate into a coalition with the aim of coordinating the effort and \emph{achieving the best tradeoff} between the performance and the associated cooperation costs''.
\subsection{Global control problem}
The overall control problem can be stated as
\begin{subequations}
\label{eq_globprob}
\begin{align}
\min_{\nu,\setE}\, & \sum_{i\in\setSP} \mathbf{J}_i(\xi_i(k),\nu_i) +  \mathbf{J}_{i}^{\chi}(\setE)\label{eq_overall_MPC} \\
& \mathrm{s.t.} \nonumber\\
& \xi_i(t+1|k)  = F_{ii}(\xi_i(t|k), \nu_i(t|k)) + \hat{\omega}_i(t|k),\label{eq_all_mod_MPC}\\
& H_i^{\mathrm{in}}(\xi_i(t|k),\nu(t|k)) \leq 0 \, t=0,\ldots,N_p, \label{eq_all_restrx_MPC}\\
& H_i^{\mathrm{eq}}(\xi(t|k),\nu(t|k)) = 0,\, t=0,\ldots,N_p-1,  \label{eq_all_restru_MPC}\\
& \xi_i(0|k) = \xi_i(k),\label{eq_all_init_constr}\\
& \setE(t) \subseteq\setN\times\setN,\, t = 0,\ldots,N_p,\label{eq_netw_constr1}\\
& \setE(t) = \setE(0),\, t = 1,\ldots,N_p,\label{eq_netw_constr2}
\end{align}
\end{subequations}
where $\setSP = \{1,\ldots,N_c\}$, and
\begin{equation}
\mathbf{J}_{i}^{\chi}(\setE)  = \sum_{t=0}^{N_p}\chi_i(t|k).
\label{eq_cost_coop_horiz}
\end{equation}
Notice that, according to constraints~\eqref{eq_netw_constr1} and~\eqref{eq_netw_constr2},  we assume the set of edges $\setE$---hence the system partition $\mathscr{P}(\setN,\mmg)$---constant during the prediction horizon $t\in[k,k+N_p]$. Problem~\eqref{eq_globprob} constitutes a dynamic optimization with mixed integer variables, which is generally not practical to solve. Since any given $\setE$ corresponds to a partition of the global system, the composition of the resulting coalitions' state and input vectors and matrices will implicitly depend on it. The choice of the network topology is made within a discrete set whose size grows exponentially with the number of subsystems.\par
\section{A game theoretical perspective}
%
The role and properties of coalitions in multi-agent interactive decision problems have been studied in game theory for decades. 
Great interest has been directed on the fair allocation of benefit among the members of a coalition. Despite the intrinsic computational complexity, which hinders its use in real-time applications, some pioneering works explored the use of coalitional game theory in engineering applications. Being natural fields for the application of game theoretic analyses, wireless networks~\cite{Saad09IEEESPM}, the smart grid~\cite{BaeyensEtAl2013, SaadEtAlSmartGrid2012} and the recharge market for plug-in electric vehicles~\cite{Malandrino2015a} have received particular attention so far.\par
\subsection{Coalition formation}
The control agents can decide with whom to cooperate and under which conditions (namely, the allocation of the payoffs derived from the cooperation). Such situation can be modeled as a coalitional game, uniquely defined by the pair $(\setN,v)$, where $\setN$ is the set of players and $v$ is the \emph{value} of a given coalition.\par
Coalition formation games consider scenarios in which the network topology and the cost for cooperation play a major role, such that the formation of a coalition is not necessarily beneficial. The theory about coalition formation games focuses on issues like: \emph{Which coalitions will form?} \emph{What is the optimal coalition size?} \emph{Which methodologies can be employed to study the properties of the resulting structures?}\par
Unlike the canonical form of coalitional games, where the fundamental assumption is that \emph{cooperation always brings benefit}, in coalition formation games gains are limited by the costs of forming a coalition. Thus, the value of the merger of two disjoint coalitions can be worse than the sum of the coalitions' separate values, that is, the \emph{superadditivity} property does not hold. Consequently, the \emph{grand coalition} (the coalition containing all the players) is seldom the optimal arrangement. Environmental changes---such as variations in the number, relevance, or constraints of the players---can affect their distribution over the coalitions. Coalition formation games can be classified in \emph{static} and \emph{dynamic}. In the first case the objective is to study the structure imposed on the coalitions by some external factor; the second category concerns the analysis of the formation of coalitions arising by the \emph{interaction} between the agents. The properties of the resulting dynamical structure and its adaptability to the environment are object of the research on dynamic coalition formation games. A monograph on this field can be found in~\cite{ray2007game}. Unfortunately, the availability of formal rules and analytical concepts is mostly limited to games in canonical form.\par
Coalition formation involves three main steps. The first two are \emph{(i)} generation of the coalition structure and \emph{(ii)} solution of the optimization problem for each coalition~\cite{SandholmEtAl1999,RamosEtAl2013}. Coalition formation is commonly studied in the form of \emph{characteristic function} games, where a value is assigned to any possible coalition $\coal \subseteq \setN$ through a function $v:2^\setN\rightarrow\mathbb{R}$. Given the graph $\mmg=(\setN, \setE)$ describing the associations among the control nodes of the system, the value of a coalition structure $\mathscr{P}(\setN,\mmg) \equiv \{\coal_1,\coal_2,\ldots,\coal_{N_c}\}$ is defined as its aggregate value
\begin{equation}
\mathscr{V}(\mathscr{P}) = \sum_{i\in \setSP}{v(\coal_i)},
\label{eq_coalstructvalue}
\end{equation}
where $\setSP= \{1,\ldots,N_c\}$.
The optimal coalition structure $\coalstr^{\ast}$ is found as the one characterized by the highest value $\mathscr{V}^{\ast}$. However, such problem has been demonstrated to be NP-complete~\cite{SandholmEtAl1999}. To overcome this issue, several solutions---resorting to heuristics, dynamic-programming, branch-and-bound algorithms---have been proposed in the literature (see~\cite{RamosEtAl2013,MAESTRE13OCAM} and references therein). Particularly interesting for control applications is the analogy first proposed in~\cite{MAESTRE13OCAM} between~\eqref{eq_coalstructvalue} and~\eqref{eq_globprob}, which serves as the foundation for a hierarchical scheme that manipulates the global controller structure (by decentralizing the feedback law over coalitions of local controllers) with regard to both the current state of the system and the communication cost. A similar architecture has been employed in the work of~\cite{FeleJPC}.\par
In characteristic form games, the value of a given coalition depends only on its members, with no regard to how the rest of the agents are organized. Such model does not apply to the vast majority of real life applications. Indeed, although games in characteristic form provide a means of modeling a wide spectrum of scenarios, it is natural in engineering applications to encounter problems in which the value of a given coalition cannot be determined regardless of how the rest of the agents are organized. Games in \emph{partition form} can model this type of problems~\cite{ChalkiadakisEtAl2012}. In these games, given a partition $\coalstr = \{\coal_1,\ldots,\coal_l\}$ of $\setN$, the value of any coalition $\coal_i\in\coalstr$ is expressed as $v(\coal_i,\coalstr)$. However, it is not possible to derive a general closed-form allocation in the considered setting. Nevertheless, in some cases the partition function game can be approximated as a characteristic function game by assigning values to coalitions following an heuristic approach: for example, if a minmax approach is employed, the value of a given coalition will take into account the most unfavorable externalities produced by any coalitional setup of the rest of agents.\par
Particularly interesting when global objectives do not take over local ones, the third and final step consists in the \emph{(iii)} distribution of the value of a coalition among its members. The \emph{payoff} $\phi_i$ is the utility received by each agent $i\in\coal$ by the division of $v(\coal)$; the vector of payoffs assigned to all the agents is referred to as the \emph{allocation}. A variety of payoff rules have been proposed in the cooperative game theory, such as the \emph{core} or the Shapley and Banzhaf values~\cite{MyersonLibroGT}. Of course, this third step is only possible if the real value $v(\coal)$ associated with coalition $\coal\subseteq\setN$ can be divided and transferred among its members (for instance, in the form of side-payments used to attract players).
Let us define $\mathbf{J}_i^{(j)}$ as the quota relative to agent $j\in\coal_i$ in the coalition cost~\eqref{eq_coal_MPC_cost}. Notice that the solution of~\eqref{eq_coal_MPC} does not imply any relation between the cost $J_j^{\ast}$ achievable \emph{independently} by any $j\in\coal_i$, and the cost $\mathbf{J}_i^{\ast (j)}$ incurred through its participation in the coalition. Provided a surplus is available, there exists a payoff assignment function such that $\mathbf{J}_i^{\ast (j)}\leq J_j^{\ast}$ is fulfilled for all $j\in\coal_i$.
As first pointed out in~\cite{AumannDreze74}, the payoff allocation concepts developed for canonical games do not admit a straightforward implementation in presence of a coalitional structure different from the grand coalition. In the same work, the definitions of the core, the Shapley value and the \emph{nucleolus} were extended to static coalition formation games, by redefining the \emph{group rationality} concept with that analogous of \emph{relative efficiency}. However, the results clearly showed that the complexity of the problem grows noticeably when  dynamic coalition formation is considered, especially when the solution has to be computed in a distributed manner~\cite{Saad09IEEESPM}. This fact motivates the application-oriented solutions found in the recent literature, such as~\cite{AptWitzel2009,ray2007game}.\par
\begin{table}[tb]
	\caption{Steps of a coalitional control algorithm.}
	\label{tab_algoritmo}
	\centering
	\begin{tabular}{ccp{10.5cm}}
		\hline
		Step & Level & Function\\  \hline
		1 & Local & Measure the state $x_i(k)$. \\[3pt]
		2 & Local & Evaluate performance $J_i$. \\[3pt]
		\multirow{5}{*}{\parbox[t]{2.5cm}{\centering 3a\\(\emph{Top-down})}} & \multirow{5}{*}{\parbox[t]{2.5cm}{\centering Global}}& Supervisory layer gathers information of the controllers' performance and evaluates alternative network topologies. Changes in the topology are allowed only if the theoretical properties of interest (for example, stability) are retained after the switching.\\[3pt]
		\multirow{4}{*}{\parbox[t]{2.5cm}{\centering 3a\\(\emph{Bottom-up})}} & \multirow{4}{*}{\parbox[t]{2.5cm}{\centering Coalition +\\neighbors}} & Collect information of the controllers involved in the coalition and its neighbors and decide whether to enable or disable links. Changes are allowed only if properties of interest are retained.\\[3pt]
		\multirow{2}{*}{3b} & \multirow{2}{*}{Coalition}  & Update the information about the structure and members of the coalition.\\[3pt]
		\multirow{3}{*}{4} & \multirow{3}{*}{Coalition} & Exchange information with the rest of the members of the coalition and calculate the control actions $\nu_i(k)$. No communication takes place with other coalitions.\\[3pt]
		5 & Local & Implement $u_i(k)$. Go to step 1.
	\end{tabular}
\end{table}
The essential steps of a coalitional control algorithm are summarized in Table~\ref{tab_algoritmo}. To handle the combinatorial explosion problem, Steps \textit{3a} and \textit{3b} can be executed at a lower rate as in~\cite{MAESTRE13OCAM}. Prior to the application of any change in the topology, theoretical properties such as stability or robustness can be checked~\cite{RiversoEtAl2014,TroddenEtAl2016ACC}. Once the structure of the coalitions is defined, the local controllers exchange information with the members of their corresponding coalitions to calculate the control actions using a distributed control scheme~\cite{MAE14DME}.
\section{Challenges for coalitional control}
As the architecture of multiagent systems becomes more complex---featuring reconfigurable topologies~\cite{FeleJPC,MAESTRE13OCAM,GrossJilgStursberg2013_ECC,NunezEtAl2015}, misaligned individual interests~\cite{FeleACC2015,BarreiroGomezEtAl2015CDC,NedicBauso2013, Malandrino2015b}, human-in-the-loop control~\cite{MaestreEtAl14CDC}, plug and play capabilities~\cite{RiversoEtAl2014}---the relationship between possible restrictions on the global availability of information and the system performance is increasingly unclear~\cite{MardenEtAl2015CDC}. Some of the challenges and open issues for coalitional control are summarized in the remainder.
\paragraph{Criteria for coalition formation}
How to determine the most appropriate coalitional structure for the overall system is a problem that shows fundamental analogies with model partitioning. Two different perspectives on coalition formation can be considered. The first is a \emph{top-down} approach, where the global coalitional structure is optimized at a supervisory layer. Of fundamental importance is the criterion used to break the overall system into coalitions. Some possibilities have been already explored, such as the minimization of an index that combines optimal control performance and communication costs~\cite{FeleJPC}, or an $H_{\infty}$ robustness index---reflecting model-plant mismatches due to nonlinearities or inaccurate identification---as proposed in~\cite{Al-GherwiEtAl2010}. To address \emph{individual rationality} as well, the second consists of a \emph{bottom-up} approach: here the formation of coalitions is produced as the outcome of an autonomous bargaining procedure. While the first approach may offer some advantages in view of global system stability guarantees, the second appeals as a more appropriate scheme for real-world (plug and play) applications.
\paragraph{Information exchange}
Information plays a fundamental role in the coordination of multiagent systems. Either attained through direct sensing or communication, information is the basis for the local decisions of agents and, as such, decisive to the emergent global behavior~\cite{MardenEtAl2015CDC}. Despite the huge effort dedicated to the development of distributed controllers for large-scale systems, communication has been mainly studied from a transmission perspective, covering issues such as bandwith limitations, data loss, or the effects of noisy channels~\cite{NairEtAlSurvey2007}. The very nature of the information exchange has received little attention. Indeed, in order to allow fundamental properties of centralized control, such as system-wide optimality and stability, the majority of the literature about distributed control overlooked privacy-related issues in order to focus on the overall system performance.\\
\emph{Does providing agents with additional information always lead to improvements in the performance? On the other hand, can the excess of information be detrimental under some given system-wide perspectives?} These questions have been partially addressed in works such as~\cite{MardenEtAl2015CDC,LangbortDelvenne2010}. In~\cite{MardenEtAl2015CDC}, a graph-coloring problem is used as a simple platform for studying the effect of information on multiagent collaboration. Results demonstrate that an increased amount of information is able to improve the efficiency of the Nash equilibria; on the other hand, it degrades the convergence rate of the distributed algorithm, where the agents seek to maximize their utility. 
\paragraph{Constraints on coalition formation}
Bigger coalitions provide better control performance at the cost of increasing cooperation requirements. The enforcement of limitations on the composition of coalitions may be necessary. Some limitations can be of direct application (as for example size constraints), whereas others can implicitly derive by the penalization of specific metrics (use of the network infrastructure, identity of the participants, optimization variables, time expected to solve the control problem). In either case, this is a problem that requires attention. See~\cite{RahwanEtAl2011} for a detailed analysis about constraints in the coalition formation process.
\paragraph{Partial cooperation and information}
When dealing with systems characterized by a strong heterogeneity, selfish interests may hinder the sharing of knowledge relevant for control purposes and cooperation as well. 
It is possible that only a subset of the control agents is willing to exchange information about their subsystems. It is therefore necessary to explore the performance bounds of the control loop possibly achievable with \emph{partial system information}. Similar issues have been already studied in the field of Economics, using tools provided by the game theory. The authors of~\cite{LangbortDelvenne2010} extend a performance metric introduced by~\cite{PapadimitriouYannakis1993}---the~\emph{competitive ratio}---for the purpose of quantifying the distance from the optimum of the distributed solution of an LP problem when the information is locally segmented. From the coalitional control standpoint, it is arguably critical to characterize the improvement provided by a broader knowledge of the system, and promote the formation of coalitions accordingly.
Notice that the previous question can be reversed: \emph{what is the minimal partition of the system model information necessary to guarantee some performance goal?} In order to address this question, the authors of~\cite{LangbortDelvenne2010} point out that additional metrics need to be formulated in order to allow the characterization of the different partitioning possibilities and the relative \emph{minimality} notion. Again, possible candidate for such task are already available in the game theory literature about multi-agent decision making under partial information.
\paragraph{Performance metrics}
The development of the coalitional control field requires the definition of key performance indicators, especially those capable of capturing the trade-off between the information exchanged by the control agents and the performance of the system. For example, in~\cite{LangbortDelvenne2010} the connection between closed-loop performance and the amount of exchanged information is characterized for distributed linear quadratic controllers. Bounds are provided on the minimal information exchange needed to achieve an improvement over the performance of the best decentralized (communication-less) scheme.\\
By addressing a networked resource allocation problem, the work of~\cite{Marden2014CDC} identifies how a measure of locality in the individual control laws can be translated into a bound on the overall achievable efficiency. 
The relationship between the redundancy of information in the control laws implemented by agents and the achievable efficiency of the overall behavior is then characterized, providing bounds on the efficiency of the stable solutions. When full information regarding the mission space is available to the agents, the efficiency of the resulting stable solutions is guaranteed to lie within 50\% of the optimal. However, as the reach of the information becomes more limited, the efficiency of the stable solutions may be as low as $1/N$ of the optimal, where $N$ designates the number of agents.
\paragraph{Commitment between cooperating agents}
Depending on the type of information exchanged and its purpose, it is possible to classify the cooperation in decreasing degrees of subsystems' integration, ranging from full information (the coalition is formally equivalent to a unique subsystem) to binding to a prearranged output interval, or sharing planned trajectories, interaction models, objectives.  
Data for different degrees of cooperation can be computed offline and retrieved during system operation, in order to implement control schemes with varying requirements (from completely decentralized to centralized) on top of the same control infrastructure~\cite{RawlingsStewart2008,FeleJPC}. This includes the possibility of reserving some backup strategies for the case of communication failure~\cite{GrossJilgStursberg2013_ECC}.
\paragraph{Theoretical guarantees}
Classical theoretical properties such as stability or robustness for the closed-loop system are more challenging in this context, specially in the case of bottom-up approaches, where little global information may be available at the individual subsystem level.
\paragraph{Coalitional guarantees}
Research on the novel properties specific of this type of control systems is needed. An example would be the degree of robustness of the coalitions with respect to the external incentives that their members may have for leaving.
\paragraph{Combinatorial explosion}
Another issue is linked with the number of possible structures of the controller under a given data network. Approaching the problem of finding the optimal coalitional structure with exhaustive search can become unviable even for relatively small number of agents, due to the exponential growth of the number of possible coalitions. Constraints on the communication topology and the number of agents in a coalition might help to relieve this issue, since the realization of some configurations would be denied. Alternatively, limitations on the number of links that can be switched during a certain period would reduce the search space and ease the on-line implementation of coalitional control schemes. 
\section{An illustrative case}
Smart grids constitute an exhaustive example of complex large-scale system. They encompass advanced power, communication, control and computing technologies. In such heterogeneous domain the game theory finds natural applications, such as energy markets and dynamic pricing. The potential of the integration of game theoretical tools in the control of smart grids is discussed in~\cite{SaadEtAlSmartGrid2012}.\par 
In existing power systems, consumers are serviced by a main electricity grid that delivers the power over the transmission lines to a substation which, in turn, delivers the power over the low-voltage distribution network. One of the building blocks of the smart grid is the microgrid, a network of distributed energy sources located at the \emph{distribution side} of the grid that can provide energy to a restricted geographical area. It may operate either together with the main grid or in an autonomous fashion. Power supply within microgrids is mainly based on small production from renewable sources or combined heat and electricity generation. Microgrids can count storage facilities (including electric vehicles' batteries) and flexible demand among their resources as well.\par
The service capacity of microgrids can be exploited to relieve the demand on the main grid. However, the intermittent generation coupled with the unpredictable nature of the demand implies that customers serviced by a microgrid may come to need extra energy from other sources. Of course, this extra energy can be provided by the main power grid. Nevertheless, the future smart grid is envisioned to encompass a large number of microgrid elements. Whenever some microgrids have an excess of production while others incur in a power shortage, a mutual exchange of energy can be beneficial for both parties, instead of relying on the main grid. The advantage of a local exchange is not limited to this: indeed, the energy transfer between nearby microgrids can significantly reduce the amount of power wasted for long-distance transmission over the distribution lines.\par
The aggregation of renewable-energy plants for their participation in the electricity market is analyzed as a canonical coalitional game in~\cite{BaeyensEtAl2013}. The main objective is the reduction of the variability in the production due to the unpredictability of renewable energy sources, exploiting the decorrelation of wind speeds at separated geographic locations. The focus is then on finding a fair mechanism for the allocation of the benefit among the producers, which can alleviate the risk associated with market participation. Intuitively, those individuals who contribute to a larger reduction in the production's variability should receive a greater share of the benefit achieved through the cooperation. The problem is solved in~\cite{BaeyensEtAl2013} through an approximated computation of the \emph{nucleolus} of the associated game, carried out as a single LP optimization that minimizes the \emph{worst-case} dissatisfaction for all possible coalitions.\par
In~\cite{SaadEtAl2011}, a cooperative energy exchange game among micro-grids, with the objective of minimizing line power losses, is formulated. The formulation involves the solution of two subproblems: an auction matching game to find the pairs of suppliers and consumers minimizing power losses, and a coalition formation game to establish the coalitions. The rules proposed in~\cite{AptWitzel2009}, based on the Pareto order, are used to decide whether to form or break coalitions. The results show that such cooperative energy exchange mechanism has the advantage of improving the autonomy of microgrids with respect to the main grid, as well as reducing the losses over the distribution lines by promoting local energy trade among neighboring microgrids. The coalition structure can be adjusted to meet variations in the demand/supply.\par
The coordination of demand and supply among communicating prosumers is addressed in~\cite{LarsenEtAl2014} with the objective of alleviating the load on substations' transformers. The focus is on the routing of information about individual demand/supply availability used for the estimation of the aggregate imbalance. The minimization of the energy imbalance of the cooperating nodes---which translates in less energy required from the main grid---is the objective of the control problem, whose distributed solution is coordinated by means of Lagrangian multipliers.\par
\begin{figure}[tb]
	\centering
	\includegraphics[width=0.65\columnwidth]{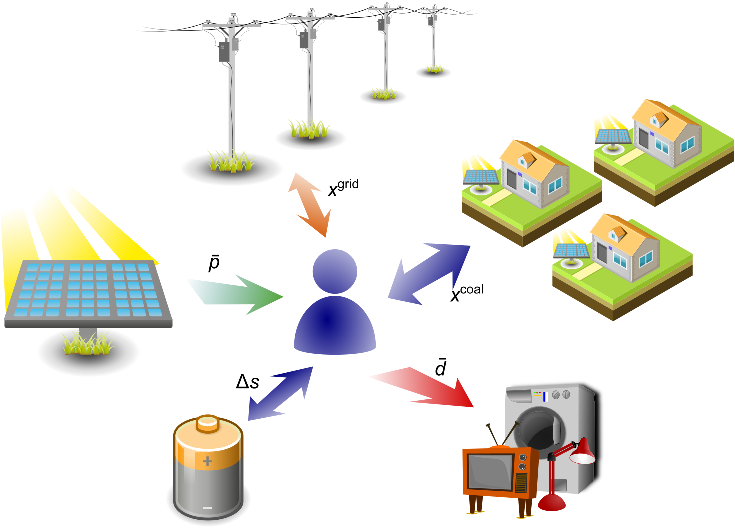}
	\caption{We consider a set of nodes connected to the main grid, equipped as well with local generation and storage devices. These prosumers can establish a local energy market by aggregating into coalitions, so as to minimize the cost of buying energy from the main grid. Power losses over the distribution lines are added to the incurred costs. As a result of the application of the coalitional control algorithm, the associations of nodes are reorganized according to variations in the local demand and supply, as well as to changes in the energy prices. The figure shows the energy balance at each node.}
	\label{fig_grandepuffo}
\end{figure}
\begin{figure}[tb]
	\centering
	\includegraphics[width=0.70\columnwidth]{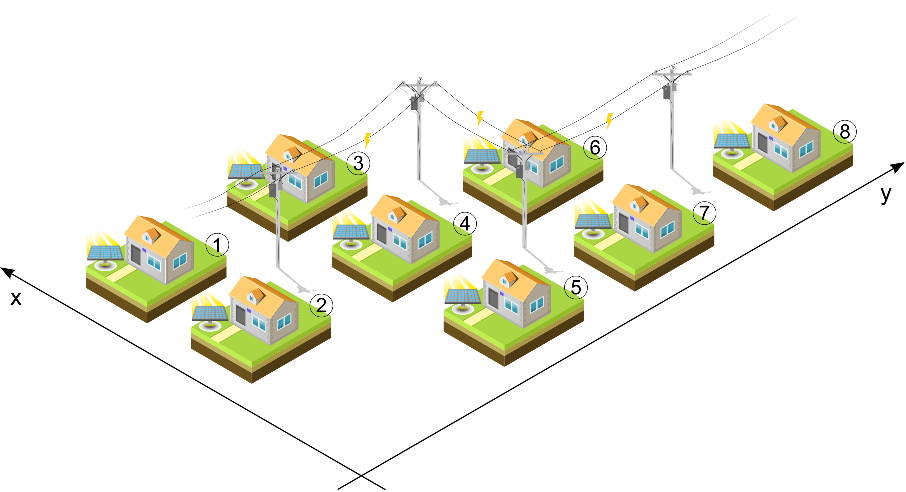}
	\caption{Reciprocal location of the 8 prosumers considered in the example: the distance between rows of houses is 0.5 km. This distance is taken into account for the losses in the distribution lines resulting from the local energy transfer.}
	\label{fig_case}
\end{figure}

Here we consider a set $\setN=\{1,\ldots,N\}$ of consumer nodes equipped with domestic production systems such as solar panels and storage devices (see Figure~\ref{fig_grandepuffo}). These nodes, situated within a small area as shown in Figure~\ref{fig_case}, pertain to the distribution layer of the grid, and are connected to the main grid through a substation.
According to the coalitional control algorithm described in the remainder, the nodes can aggregate into coalitions. Inside each coalition, the power transfer is the result of an optimal control problem, formulated such as to minimize the cost of buying energy from the main grid \emph{for the entire coalition}. Power losses over the distribution lines due to the energy transfer between members of the coalition are added to the incurred costs (costs of power losses from the main grid are not considered separately from the spot price). It is interesting to see how, as a result of the application of the coalitional control algorithm, the associations of nodes are reorganized according to the variations in the local demand and production, as well as to the changes in the energy prices.\par
Let us consider first the case in which the consumer nodes only rely on the main grid. For any node $i\in\setN$, the energy exchange with the main grid is quantified by the variable $\xred_i$, representing the difference $\bar{d}_i-\bar{p}_i$ between local demand and generation. Whenever $\xred_i>0$, node $i$ buys energy $d_i\red = \xred_i$ from the grid---to fulfill the residual demand that cannot be matched by its local production---at the spot price $c_i\red$. In case of local energy surplus instead, that is $\xred_i<0$, node $i$ can sell energy $p_i\red$ to the grid at the feed-in tariff $v_i\red$. We assume that both generation and demand profiles are deterministic, known beforehand within a 5-hour horizon. We also assume that the tariff scheme implemented by utility companies is always such that $c_i\red(k)>v_i\red(k)$. The local availability of a storage device allows the agent some flexibility when facing the spot prices imposed by the grid: the purchase and the sale of energy can be shifted in order to benefit of more convenient prices. At this point, the control problem individually addressed by each node would be (symbols are defined in Table~\ref{tab_ex_symbol})
\begin{table}[tb]
	\caption{List of symbols employed in the description of the example.}
	\label{tab_ex_symbol}
	\centering
	\begin{tabular}{ l l l }
		Symbol & Description & Unit \\ \hline\\[-1.5ex]
		$\bar{d}$ & Demand & [kWh] \\ 
		$\bar{p}$ & Generation output  & [kWh] \\  
		$d\red$ & Energy bought from the grid  & [kWh] \\ 
		$p\red$ & Energy sold to the grid  & [kWh] \\ 
		$c\red$ & Grid spot price & [CU] \\
		$v\red$ & Feed-in tariff & [CU] \\
		$d\excoal$ & Energy bought from the coalition  & [kWh] \\ 
		$p\excoal$ & Energy sold to the coalition  & [kWh] \\
		$c\excoal$ & Coalition spot price & [CU] \\
		$v\excoal$ & Coalition feed-in tariff & [CU] \\
		$s$ & State of charge of the storage & [kWh] \\  
		$\Delta s$ & Increment in the storage level & [kWh] 
	\end{tabular}
\end{table}
\begin{subequations}
\label{eq_ind_example}
\begin{align}
\min_{\bfu_i}\, & \sum_{t=0}^{N_p-1} c_i\red(t|k) d_i\red(t|k) - v_i\red(t|k)p_i\red(t|k), \label{eq_ind_ex_cost}\\
& \mathrm{s.t.} \nonumber\\
& s_i(t+1|k) = s_i(t|k) + \Delta s_i(t|k),\label{eq_ind_ex_stor}\\
& \bar{d}_i(t|k) + \Delta s_i(t|k) + p_i\red(t|k) = \bar{p}_i(t|k) + d_i\red(t|k),\label{eq_ind_ex_balan}\\
& d_i\red(t|k),p_i\red(t|k) \geq 0,\label{eq_ind_ex_restru}\\
& s_i(t|k) \in [0,s_{i,\max}],\,t = 0,\ldots,N_p,\label{eq_ind_ex_restrstor}\\
& s_i(0|k) = s_i(k).\label{eq_ind_ex_init_constr}
\end{align}
\end{subequations}
The manipulable inputs are $\bfu_i \triangleq [\Delta\mathbf{s}_i\trasp, \mathbf{d}_i\trasp,\mathbf{p}_i\trasp]\trasp$,
where the bold notation indicates a vector containing the values along the prediction horizon.
The monetary amount expressed by~\eqref{eq_ind_ex_cost} is the local balance of the energy exchange with the grid. Since the problem is formulated as a minimization, a positive balance indicates the expense required to buy energy from the grid, whereas a negative balance is the revenue from feeding energy to the grid. Constraints~\eqref{eq_ind_ex_stor}--\eqref{eq_ind_ex_init_constr} represent, respectively, the dynamics of the storage, the energy balance at the node, the admitted values of inputs and storage level, and the initial state of charge.\par
Problem~\eqref{eq_ind_example} is now modified taking into account the possibility for two or more nodes to agree over mutual exchange of energy, establishing a common pool of resources. The local variable $\xcoal_i \triangleq d\excoal_i - p\excoal_i$ , whose components are the demand of energy to the coalition and the energy transferred to its members, describes such exchange. The objective function of the coalitional problem involves a component for the minimization of the costs of energy exchange with the main grid, and an additional component concerning the minimization of power losses due to transfers between members of the coalition.
\begin{subequations}
\label{eq_coal_example}
\begin{align}
\min_{\boldsymbol{\nu},\coal} \, & \rho_{\mathrm{coal}} \sum_{i,j\in\coal} \sum_{t=0}^{N_p-1} r_{ij}\left(\frac{\xcoal_{ij}(t|k)}{2}\right)^2 + \sum_{i\in\coal} \sum_{t=0}^{N_p-1}  c_i\red(t|k) d_i\red(t|k) - v_i\red(t|k)p_i\red(t|k), \label{eq_coal_ex_cost}\\
& \mathrm{s.t.} \nonumber\\
& s_i(t+1|k) = s_i(t|k) + \Delta s_i(t|k),\, i\in\coal\label{eq_coal_ex_stor}\\
& \bar{d}_i(t|k) + \Delta s_i(t|k) + p_i\red(t|k) + p_i\excoal(t|k) = \bar{p}_i(t|k) + d_i\red(t|k) + d_i\excoal(t|k),\, i\in\coal,\label{eq_coal_ex_balan}\\
& \sum_{i\in\coal} p_i\excoal(t|k) = \sum_{i\in\coal} d_i\excoal(t|k),\label{eq_coal_ex_cbalan}\\
& d_i\red(t|k),p_i\red(t|k) \geq 0,\,i\in\coal,\label{eq_coal_ex_restru}\\
& s_i(t|k) \in [0,s_{i,\max}],\,t = 0,\ldots,N_p,\, i\in\coal\label{eq_coal_ex_restrstor}\\
& s_i(0|k) = s_i(k),\, i\in\coal,\label{eq_coal_ex_init_constr}\\
& \coal\subseteq\setN,\label{eq_coal_ex_coal_constr1}
\end{align}
\end{subequations}
where $\xcoal_{ij}$ is the interchange of energy among the pair of agents $i,j\in\coal$, and $r_{ij}$ is their relative Euclidean distance. The coalitional energy transfer $d_i\excoal-p_i\excoal$ is taken into account in the balance constraints~\eqref{eq_coal_ex_balan}. Constraint~\eqref{eq_coal_ex_cbalan} express the zero sum energy balance of the coalition.\par
Given its complexity, problem~\eqref{eq_coal_example} is approximated here by splitting it in two subproblems, each one concerning one of the two components of the objective function. 
So, for any $\coal\subseteq\setN$, the first subproblem problem is formulated as:
\begin{equation}
\min_{\boldsymbol{\nu}} \mathbf{J} = \sum_{i\in\coal} \sum_{t=0}^{N_p-1}  c_i\red(t|k) d_i\red(t|k)  - v_i\red(t|k)p_i\red(t|k), 
\label{eq_coal1_ex_cost}
\end{equation}
subject to~\eqref{eq_coal_ex_stor}--\eqref{eq_coal_ex_init_constr}. The resulting energy transfers are accompanied by an energy loss over the distribution lines: an approximation of these is added to $\mathbf{J}$ as
\begin{equation}
\mathbf{J}^{\chi} = \rho_{\mathrm{coal}} \hat{r} \sum_{i\in\coal} (d_i\excoal)^2,
\label{eq_coal_powerloss}
\end{equation}
where $\hat{r}$ is a representative mean distance between every pair of coalition members, and $d_i\excoal$ is the energy demanded to the coalition by agent $i$. Finally, the value $v(\coal)$ of any given coalition $\coal\subseteq\setN$ (recall that $v:2^{\setN}\mapsto \mathbb{R}$ is the function mapping any possible coalition of agents into a real value) is defined as $v(\coal) \triangleq \mathbf{J} + \mathbf{J}^{\chi}$, where~\eqref{eq_coal_powerloss} represents the cost of forming the coalition.\par
Now, the second subproblem consists in finding the partition $\coalstr = \{\coal_1,\ldots,\coal_{N_c}\}$ of $\setN$ such that the preference of every agent is satisfied according to the Pareto order. In order to do this, individual payoffs within each possible subset of agents $\setS\subseteq\setN$ are first calculated as those defined by the Shapley value $\phi^{\setS}: \mathbb{R}^{2^{|\setS|}}\mapsto \mathbb{R}^{|\setS|}$. Thus, for $i\in\setS$:
\begin{equation}
\phi_i^{\setS}(v) = \sum_{\coal\subseteq\setS\setminus\{i\}}\frac{\left|\coal\right|!(|\setS|-|\coal|-1)!}{|\setS|!}\left[v(\coal\cup\{i\})-v(\coal)\right].
\label{eq_Shapley}
\end{equation}
Note how the marginal contribution $v(\coal\cup\{i\})-v(\coal)$ is weighted by the probability for any agent $i$ of joining the coalition $\coal\subseteq\setS\setminus\{i\}$, in case the agents form the coalition $\setS$ in a random order: in this way, the value assigned by Shapley's criterion corresponds to the individual \emph{expected} marginal contribution.\par
Once this step is accomplished, a mapping $\Phi: \setN\times 2^{\setN} \mapsto \mathbb{R}$ of the individual payoff for each possible coalition the agent  $i\in\setN$ can participate in is available. 
At time step $k$, $\Phi(i,\coal,k)$ defines the cost (or the benefit) incurred by agent $i$ by participating in the coalition $\coal\subseteq\setN$. Note that such payoff will coincide with the cost of energy exchange with the main grid ($\Phi(i,\{i\},k) \equiv c_i\red(k) d_i\red(k)  - v_i\red(k)p_i\red(k)$) in case the agent does not participate in any coalition. On the other hand, if the agent is member of a coalition, the \emph{equivalent} price payed/earned on the coalition's internal market can be derived as:
\begin{equation}
c_i^{\mathrm{eq}}(k)\triangleq \frac{\Phi(i,\coal,k)}{x_i\red(k) + x_i\excoal(k)}\,\left[\frac{\mathrm{CU}}{\mathrm{kWh}}\right],
\label{eq_coalprice}
\end{equation}
where CU stands for currency unit.
Such mapping provides each agent with a preference order over the coalitions he wishes to join. Since any agent can be in one coalition at a time (coalitions do not overlap), the agents will organize themselves into a partition $\coalstr^{\ast}$ \emph{following the preference order dictated by} $\Phi$.\par
\begin{figure}[tbp]
	\centering
	\subfloat[$\rho_{\mathrm{coal}}=5\cdot 10^{-3}$]{%
		\includegraphics[width=0.70\columnwidth]{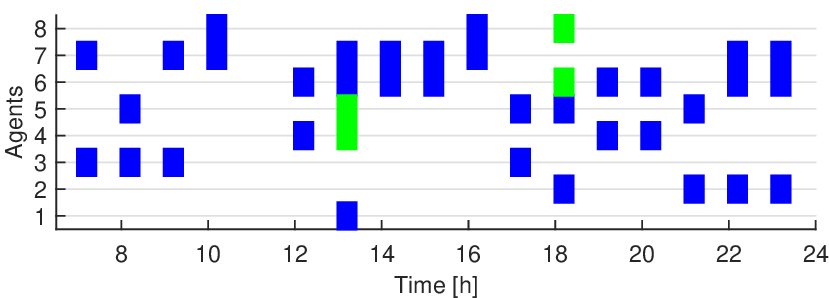}
	}
	
	\subfloat[$\rho_{\mathrm{coal}}=5\cdot 10^{-4}$]{%
		\includegraphics[width=0.70\columnwidth]{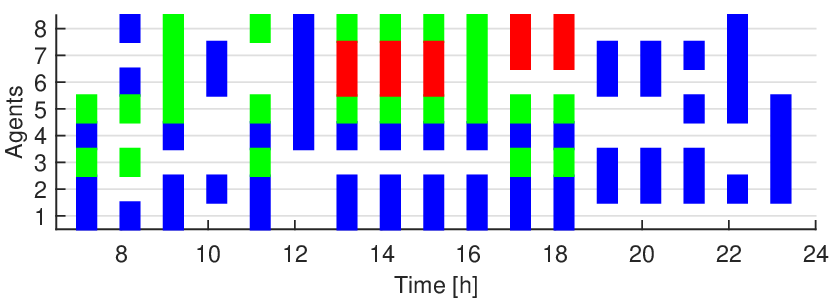}
	}
	
	\subfloat[$\rho_{\mathrm{coal}}=10^{-4}$]{%
		\includegraphics[width=0.70\columnwidth]{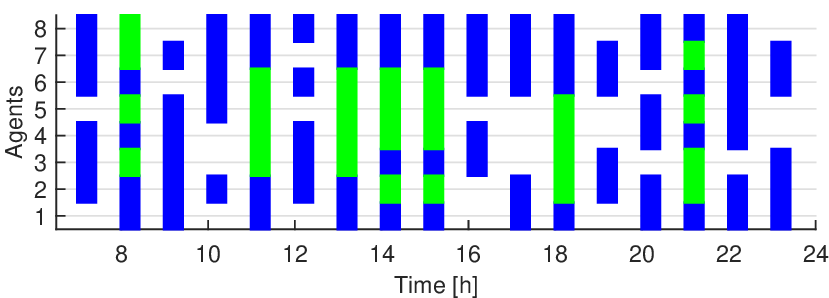}
	}
	
	\subfloat[$\rho_{\mathrm{coal}}=10^{-5}$]{%
		\includegraphics[width=0.70\columnwidth]{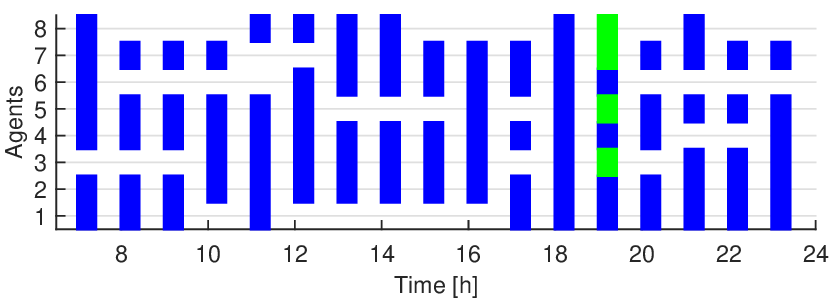}
		\label{fig_coals_1e-5}
	}
	\caption{Evolution of the coalitions among the nodes, under different penalties $\rho_{\mathrm{coal}}$ for energy losses. Agents are allowed to reevaluate their affiliation at each time step, by updating the preference mapping $\Phi(i,\cdot,k)$. When costs for local energy transfers are significant, coalitions typically involve a restricted number of neighboring agents. Conversely, big cooperating clusters form when penalties for power losses are low.}
	\label{fig_coals}
\end{figure}
\begin{figure}[tbp]
	\centering
	\includegraphics[width=0.65\columnwidth]{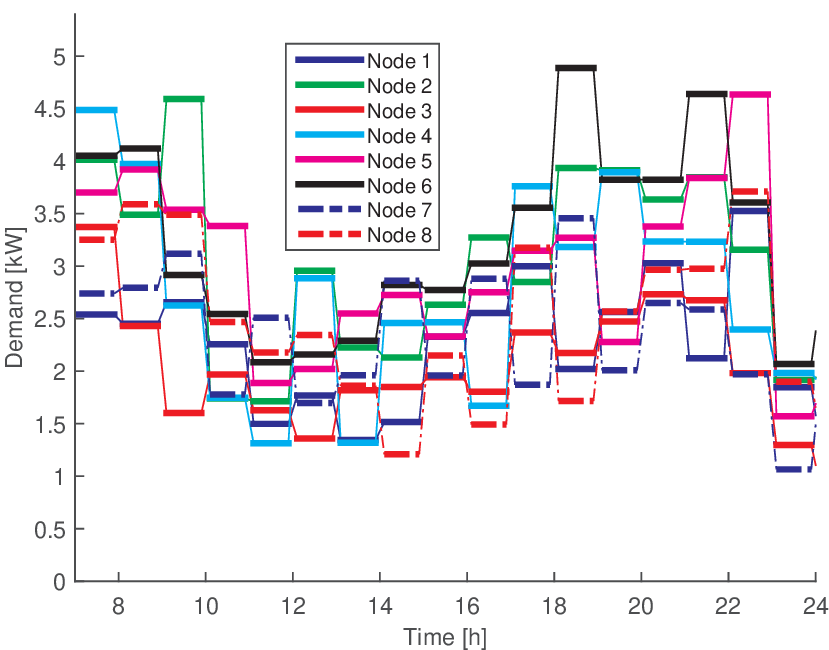}
	\caption{Daily demand patterns considered in the simulations for the 8 nodes.}
	\label{fig_demand}
\end{figure}
\begin{figure}[tbp]
	\centering
	\includegraphics[width=0.65\columnwidth]{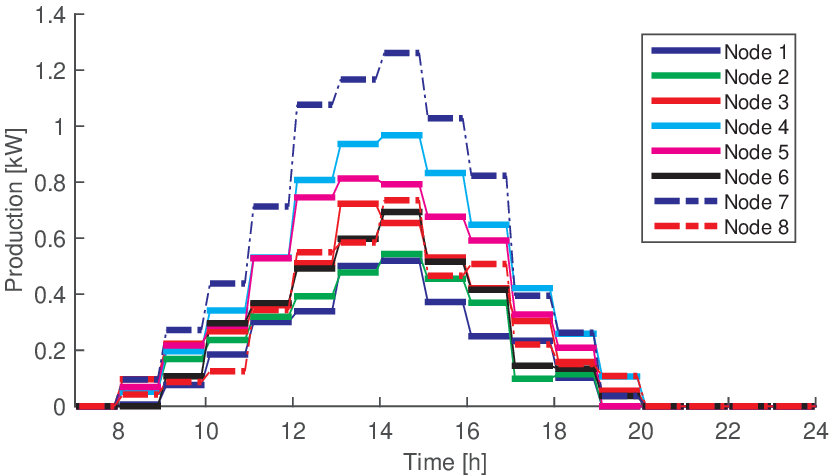}
	\caption{Generation profiles considered in the simulations for the 8 nodes.}
	\label{fig_generation}
\end{figure}
\begin{figure}[tbp]
	\centering
	\includegraphics[width=0.65\columnwidth]{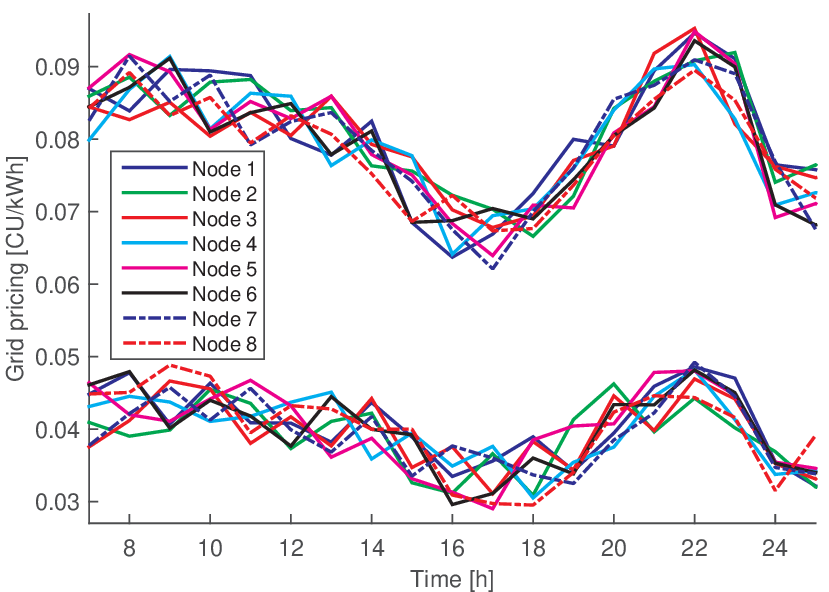}
	\caption{Grid spot prices relative to the energy purchase (top) and excess of production sale (bottom).}
	\label{fig_prezzi}
\end{figure}

The demand and generation patterns considered in the simulation are depicted in Figures~\ref{fig_demand} and~\ref{fig_generation}. Grid spot prices are shown in Figure~\ref{fig_prezzi}, relative to the purchase (top) and feed-in (bottom). The interval 7 a.m.--24 p.m.~is considered in the simulations. A time step of 1 h and a prediction horizon $N_p = 5$ h are employed in the optimization problem. We assume the possibility for each node to access the grid with different prices, to emulate the presence of multiple utility companies. Notice that this means that coalitions do not merely consist of a combination of nodes with a surplus of power and nodes that are in need of additional power to meet their demand. Indeed, nodes can make agreements to get access to the main grid through the more advantageous tariff of another node. The evolution of the coalitions, under different penalties $\rho_{\mathrm{coal}}$ for energy losses, is shown in Figure~\ref{fig_coals}. By recomputing the mapping $\Phi(i,\cdot,k)$, agents can reconsider their affiliation at each time step. When costs for local energy transfers are significant, coalitions typically involve a restricted number of neighboring agents. On the other hand, big cooperating clusters form when penalties for power losses are low.


\begin{figure}[tbp]
	\centering
	\includegraphics[width=0.65\columnwidth]{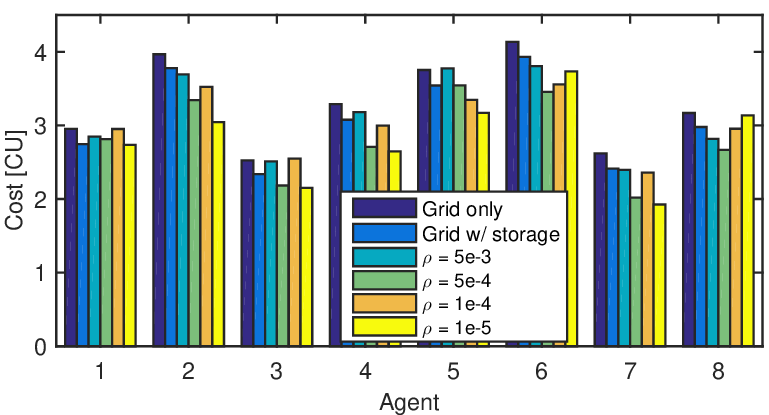}
	\caption{Comparison of the costs incurred by each agent during the simulated interval, over different scenarios: \emph{(i)} no coalitions, no storage \emph{(ii)} local storage available \emph{(iii)} coalitions with different costs for power losses. Penalties for energy losses influence the final cost as expected. Notice that the preference order over joining the coalitions is formulated over a given time horizon. However, agents are allowed to change their affiliation at each time step. This may cause the accumulation of some short term monetary losses, visible in the results.}
	\label{fig_costi}
\end{figure}
The costs incurred by the agents for the whole simulated interval are shown in Figure~\ref{fig_costi}. As expected, internal transfers' power losses affects the final result. The reason for the high costs for agents 6 and 8 with $\rho_{\mathrm{coal}}=10^{-5}$ can be ascribed to the fact that they are mostly left out of coalitions, as can be seen in Figure~\ref{fig_coals_1e-5}. Furthermore, recall that the payoff mapping, and thus the decision whether to join a coalition, is based on a forecast along the horizon $N_p = 5$ h (see~\eqref{eq_coal1_ex_cost}). However, since agents are allowed to join/leave coalitions every time step, possible monetary losses taking place in the short term may accumulate (this issue is pointed out in~\cite{BaeyensEtAl2013} as well).
The realizations of the equivalent prices during the simulations are shown in Table~\ref{tab_eqprices}. Comparing the average prices resulting by the coalitions' internal market with the grid tariffs in Figure~\ref{fig_prezzi}, it is clear that the Shapley allocation benefits both buyers and sellers. Buyers can access energy at prices more affordable than those offered by the main grid, while sellers can achieve an higher profit on the local market.\par
\begin{table}[tb]
	\caption{Comparison of the average prices (in [CU/kWh]) paid by each agent. The figures in the third column are the equivalent average prices relative to the coalitions' internal market. See how, in reference to the tariffs in Figure~\ref{fig_prezzi}, both buyers and sellers are benefited by the local energy market. Demand can be satisfied at prices lower than those offered by the main grid, and supply gets better value on the local market.}
	\label{tab_eqprices}
	\centering
	\begin{tabular}{ c c c c }
		\multirow{2}{*}{Agent} & Grid only & Grid only & Coalitions \\ 
		& no storage & w/ storage & ($\rho = 10^{-5}$) \\\hline\\[-1.5ex]
		1 & 0.0817 & 0.0787 & 0.0784 \\
		2 & 0.0813 & 0.0795 & 0.0603 \\
		3 & 0.0810 & 0.0782 & 0.0653 \\
		4 & 0.0812 & 0.0784 & 0.0620 \\
		5 & 0.0822 & 0.0797 & 0.0714  \\
		6 & 0.0805 & 0.0784 & 0.0745 \\
		7 & 0.0810 & 0.0776 & 0.0620 \\
		8 & 0.0810 & 0.0786 & 0.0798
	\end{tabular}
\end{table} 
In conclusion, local energy trade has been induced among consumers within a small geographical areas through the implementation of a coalitional control scheme. Power losses over the distribution lines can be taken into account in the coalition formation mechanism. Naturally, the larger the number of agents participating in such scheme, the higher the chance to match demand and supply among them, resulting in a higher profitability of the coalitions.

\section{Conclusion}

The advances in information and communication technologies point towards a future where the number of networked entities grows larger and larger: hence, there is a need for promoting their coordination when it becomes relevant in terms of costs or performance. Nevertheless, these entities may have incentives for cooperating only to a limited extent or keeping part of their information private. Also, the possibility for the entities to join and leave the network in a plug and play fashion is foreseen as a feature of future infrastructures. 
The effort made within the distributed control engineering community to formulate answers to such challenging issues will surely benefit from the insights provided by the cooperative game theory. In this regard, coalitional control represents the natural evolution of distributed control and aims at providing flexible schemes---balancing the overall performance and the coordination effort---featuring mechanisms and incentives to enable the cooperation in a way that all the entities can fulfill their own constraints and objectives.

\section{Acknowledgments}
The authors are grateful to the anonymous referees and the Editor for helpful suggestions and comments. Finally, the authors would like to thank Prof.~Daniel Lim\'on Marruedo, Ezequiel Gonz\'alez Debada and Pablo B\'aez Gonz\'alez for the enlightening discussions and their help.



\newpage

\bibliographystyle{IEEEtran}
\bibliography{biblioCSM}

\end{document}